\documentclass[twocolumn,showpacs,preprintnumbers,amsmath,amssymb]{revtex4}

\usepackage{bm}

\begin{document}

\bibliographystyle{unsrt}

\preprint{math-ph/0307038}

\title{Physical Space as a Quaternion Structure, I:\\Maxwell Equations.
A Brief Note.}

\author{Peter Michael Jack}
 \altaffiliation[Alumnus of ]{the Physics Department of Columbia
 University, NY.} 

\affiliation{%
Hypercomplex Systems\\
Toronto, Canada
}%
\email{math@hypercomplex.com}

\date{July 18, 2003}

\begin{abstract}
This paper shows how to write Maxwell's Equations in Hamilton's
Quaternions. The  fact that the quaternion product is non-commuting
leads to distinct left and right derivatives which must both be
included  in the theory. A new field component is then revealed, 
which reduces part of the degree of freedom found in the gauge, but 
which can then be used to explain thermoelectricity, suggesting that 
the theory of heat has just as fundamental a connection to
electromagnetism as the magnetic field has to the electric field, 
for the new theory now links thermal, electric, and magnetic
phenomena  altogether in one set of elementary equations. This
result  is based on an initial hypothesis, named ``The Quaternion
Axiom,''  that postulates physical space is a quaternion structure. 
\end{abstract}

 

\maketitle

\section{\label{sec:level1}THE QUATERNION AXIOM.\protect\\}

The quaternions 
\begin{equation}
a = a0 + a1.i + a2.j + a3.k
\end{equation}

where i, j, k are the anti-commuting hypercomplex roots of -1, 
and the a0, a1, a2, a3 are elements of the real number set, can 
be used to write down Maxwell's Equations. We postulate that 
physical space is a quaternion structure, so that the units \{\ i, j,
k\}\ 
 represent space dimensions, while the scalar \{\ 1\}\
represents time, and the space units obey the product rules 
given by W. R. Hamilton in 1843\footnote{W. R. Hamilton, 1844, 
{\it On a new species of Imaginary Quantities connected with the 
           Theory of Quaternions} [communicated November 13, 1843],
Ir. Acad. Proc., II, 424-434.} ; 

\begin{eqnarray}
& i^2 = j^2 = k^2 = -1 \nonumber
\\
\\
	& i = jk = -kj,  j = ki = -ik,  k = ij = -ji \nonumber
\end{eqnarray}

We shall refer to this postulate as The Quaternion Axiom. A 
position vector in this quaternion space will take the form, 
$r = ct + ix + jy + kz$, with c the characteristic speed that 
links clock ticks to metre rules, equivalent here to the speed 
of light, so that all measures within this quaternion 4-vector 
are ultimately in the same units-of-length. 

\subsection{\label{sec:level2}Ambiguous Product.\protect\\}

Now, because quaternions don't commute, we need to recognise 
the distinction between left and right actions. Consider two 
quaternion variables a \& b. Let us define the 
symbols `$\rightarrow$' and `$\leftarrow$' to mean `operate to the right' and 
`operate to the left', respectively. Then, for example, if the 
`a' term is the operator, and the `b' term is the variable acted 
upon by the operator, we have; 

\begin{eqnarray}
  a\rightarrow b  & = & a0.b0 - a1.b1 - a2.b2 - a3.b3  \nonumber
\\
	& + & a0.(b1.i + b2.j + b3.k) \nonumber
\\
	& + & (a1.i + a2.j + a3.k).b0
\\
	& + & (a2.b3 - a3.b2).i \nonumber
\\
        & + & (a3.b1 - a1.b3).j \nonumber
\\
        & + & (a1.b2 - a2.b1).k \nonumber
\end{eqnarray}
\begin{eqnarray}
  b\leftarrow a  & = & a0.b0 - a1.b1 - a2.b2 - a3.b3 \nonumber
\\
	& + & a0.(b1.i + b2.j + b3.k) \nonumber
\\
	& + & (a1.i + a2.j + a3.k).b0
\\
	& - & (a2.b3 - a3.b2).i \nonumber
\\
        & - & (a3.b1 - a1.b3).j \nonumber
\\
        & - & (a1.b2 - a2.b1).k \nonumber
\end{eqnarray}

Note, we only need the symbols, $\rightarrow$ and $\leftarrow$\ ,
between the quaternions themselves. Once we have resolved the algebra
to the  level of the components, we can revert to the usual
convention of  putting the operator on the left and the variable
acted upon  on the right. 

If there is no physical reason to select one product over the other, 
both products must be accounted for equally in the expressions used 
to model phenomena in any given theory, otherwise such expressions 
will have an inherent left-hand or right-hand bias.

When dealing with operator products therefore, we define the right 
product by, $a\rightarrow b$, and the left product by, $b\leftarrow
a$.  Then the two combinations of these, the symmetric product,
\{a,b\},  and the antisymmetric product, [a,b], are defined 
correspondingly by; 
\begin{equation}
	\{a,b\} =  (1/2)(a\rightarrow b + b\leftarrow a)
\end{equation}
\begin{equation}
	[a,b] =  (1/2)(a\rightarrow b - b\leftarrow a)
\end{equation}

Which product we employ in our theory, and where, is dictated by the
symmetries  inherent in the physical problem.

\section{\label{sec:level1}MAXWELL EQUATIONS.\protect\\}

Now, let the Electromagnetic Potential be 
\begin{equation}
	A = U + A1.i + A2.j + A3.k 
\end{equation}

and the differential operator $(d/dr)$ be defined by, 
\begin{equation}
	\frac{d}{dr} = \frac{1}{c} \frac{\partial }{\partial t} 
        + \frac{\partial }{\partial x}\ i
        + \frac{\partial }{\partial y}\ j
        + \frac{\partial }{\partial z}\ k
\end{equation}

then by inspection, the Electric and Magnetic fields as quaternions are, 
\begin{equation}
	E = -\{d/dr,A\}  = -(1/2)(d/dr\rightarrow A + A\leftarrow d/dr)
\end{equation}
\begin{equation}
	B = +[d/dr,A]  = +(1/2)(d/dr\rightarrow A - A\leftarrow d/dr)
\end{equation}

That is, the electric field is the negative symmetric derivative of
the  potential, and the magnetic field is the positive antisymmetric 
derivative of the potential. The space components of these quaternion
fields correspond exactly to the electric and magnetic fields in the
usual 3-vector calculus. However, the electric quaternion field now
has a time component, which we label, T, so that, $E = T + \mathbf E$,
while the magnetic quaternion field has no time component, so that, 
$B = 0 + \mathbf B$. And if we allow our notation to alternate between 
Heaviside-Gibbs 3-vector and that of Hamilton's Quaternion 3-vector, 
taking care to match up only the components of the appropriate
expressions, we can write the quaternion derivative in terms of the 
more familiar vector notation, 

\begin{equation}
    \frac{d}{dr}\rightarrow A = \frac{1}{c} \frac{\partial
    U}{\partial t} - div(\mathbf A) + \frac{1}{c} \frac{\partial \mathbf A}{\partial
    t} + grad(U) + curl(\mathbf A)
\end{equation}

\begin{equation}
    A\leftarrow \frac{d}{dr} = \frac{1}{c} \frac{\partial U}{\partial
    t} - div(\mathbf A) + \frac{1}{c} \frac{\partial \mathbf A}{\partial
    t} + grad(U) - curl(\mathbf A)
\end{equation}
\\
where we observe that the quaternion derivative has ``five'' distinct
parts, much like the  five fingers on the human hand. The electric
field is then obtained  from clapping the hands together, so that the
distinction between  the two vanish (and consequently giving rise, as
we  shall see, to ``heat'' as well as electricity). While, the magnetic
field is obtained from the opposite action of pulling the hands
apart, so that the distinction between the two become more pronounced. 
\\

Now, although we could invent constructions using the quaternion
conjugate to eliminate the time component from all our calculations
and so obtain the exact results of the 3-vector calculus, this would 
make the calculations more contrived and less natural than the simple
structure presented here. We therefore let the natural structure of 
the algebra lead us instead, to the results that follow from
accepting this new field component, rather than try to eliminate it 
from our framework. Then, by inspection, the reformulated
{\bf Maxwell Field Equations} are,

\begin{equation}
	[d/dr,B] = +\{d/dr,E\}
\end{equation}
\begin{equation}
	[d/dr,E] = -\{d/dr,B\}
\end{equation}

The antisymmetric derivative of the magnetic field is the positive of
the symmetric derivative of the electric field. And, the
antisymmetric derivative of the electric field is the negative of the
symmetric derivative of the magnetic field. The first represents a
real physical law, while the second is easily proven to be an algebraic 
identity when given the definitions of the electric and magnetic
fields above. 
\\

When written in the vector notation of Heaviside-Gibbs, these two
4-vector (quaternion) equations become the usual four 3-vector equations, 

\begin{eqnarray}
curl(\mathbf B) & = & +1/c.\partial \mathbf E/\partial t + grad(T) \\
curl(\mathbf E) & = & -1/c.\partial \mathbf B/\partial t \\
div (\mathbf E) & = & +1/c.\partial T/\partial t \\
div (\mathbf B) & = & 0
\end{eqnarray}
\begin{eqnarray}
T & = & -1/c.\partial U/\partial t + div(\mathbf A) \\
\mathbf E & = & -grad(U) - 1/c.\partial \mathbf A/\partial t \\
\mathbf B & = & curl(\mathbf A)
\end{eqnarray}
\\
provided we now identify the electric charge density, $\rho$, and
electric current density,  $\mathbf J$, with the terms involving T. Thus, 
$4\pi \rho = +1/c.\partial T/\partial t$, and $4\pi \mathbf J/c =
grad(T)$.

The scalar
quantity,  T, we shall call the ``Temporal Field.'' This scalar 
field has the same units-of-dimension as the electric and magnetic 
vector fields in our Gaussian system of units. And thus for a given 
charge, q, the quantity, qT, has units of force, similar to the 
electric force, $q\mathbf E$, and the magnetic force, $q\mathbf v/c
\times \mathbf B$. However, this 
scalar force has no space direction. Instead, it acts along the 
timeline, since that is the scalar axis under The Quaternion Axiom 
used as the basis of this derivation. 
\\

What effect is produced by a scalar force acting along the timeline? 
\section{\label{sec:level1}THERMOELECTRICITY.\protect\\}

The electric force moving a charge, q, through a displacement,
$d\mathbf x$,  does work, $dW = q\mathbf E \mathbf \cdot d\mathbf x$. And work is a form of energy. This energy
is  delivered to the external environment, and is available, for
example, as mechanical work, capable of moving the parts of a mechanical 
system against a frictional resistance. 
\subsection{\label{sec:level2}Heat.\protect\\}
Similarly, the temporal force
acting on a charge, q, for a time interval, cdt, produces an
energy-like  term, $dW = -qTcdt$ (the minus sign reflecting the
opposite signs in the squares of the unit time, $1^2 = +1$, and unit
space, $i^2 = j^2 = k^2 = -1$, measures, in the quaternion spacetime,
needed here because $\mathbf E$ and $d\mathbf x$ are Heaviside-Gibbs vectors; thus a
positive charge, $q > 0$, under the influence of a positive value
temporal field, $T > 0$, produces the equivalent of negative work, i.e.
the charge-field interacting system will absorb energy from its
surroundings, positive charges thus effectively appearing ``cold,'' 
while negative charges effectively appearing ``hot''). 
\\

In this case, 
over the given time interval, energy is absorbed or evolved from the 
charge-field interacting system accordingly as the signs of the
charge and the temporal field are the same or opposite. Since this 
scalar energy does not require the charge to move in space, in order 
to materialize as some observable physical phenomena the energy that
is absorbed and/or evolved must manifest as a form of heat. Moreover,
this heat is proportional to the first power of the charge, and thus 
reverses sign with the change in sign of the charge, or
correspondingly a change in sign of the electric current, making this
a reversible heat, corresponding to the experimental observations
already known as Peltier and Thomson Heats in thermoelectricity. 
\\

Thus, the temporal field, 
$T = -1/c.\partial U/\partial t + div(\mathbf A)$, represents the total heat energy per unit 
charge evolving per unit time (i.e. time measured in units of length)
from the charge-field interacting system due to both the loss in 
electrostatic potential energy at the location and the flow of 
electrodynamic momentum out of the same location, much like the 
diffusion heat flow being due to the sum of two different processes, 
conduction and convection, in nonequilibrium thermodynamics. 
\\

Now, instead of using the symmetric and antisymmetric derivatives,
the two quaternion electromagnetic equations can also be written
using the left and right derivatives of the potential. 
\begin{eqnarray}
d/dr\rightarrow (d/dr\rightarrow A) & +\ (A\leftarrow d/dr)\leftarrow
d/dr & =  0 \ \ \ \  \\
d/dr\rightarrow (A\leftarrow d/dr) & -\ (d/dr\rightarrow A)\leftarrow
d/dr & =  0 \ \ \ \ 
\end{eqnarray}

When we introduce the electric charge density and electric current
source terms, as additional inhomogeneous parameters to the temporal
terms, instead of identifying the electric source directly with the 
temporal terms themselves, the second equation is unchanged, and the 
first equation says, 
\begin{equation}
d/dr\rightarrow (d/dr\rightarrow A) + (A\leftarrow d/dr)\leftarrow
d/dr = 8\pi J \ \ \ 
\end{equation}
\\
where, $J = (\rho, \mathbf J/c)$. In the Heaviside-Gibbs 3-vector
format, the new inhomogeneous electromagnetic equations therefore become, 
\begin{eqnarray}
curl(\mathbf B)  & = & +1/c.\partial \mathbf E/\partial t + grad(T) +
4\pi \mathbf J/c \\
curl(\mathbf E)  & = & -1/c.\partial \mathbf B/\partial t \\
div (\mathbf E)  & = & +1/c.\partial T/\partial t + 4\pi \rho \\
div (\mathbf B)  & = & 0
\end{eqnarray}
\begin{eqnarray}
T & = & -1/c.\partial U/\partial t + div(\mathbf A) \\
\mathbf E & = & -grad(U) - 1/c.\partial \mathbf A/\partial t \\
\mathbf B & = & curl(\mathbf A)  
\end{eqnarray}

These equations then make a clearer distinction between the ``thermal'' and
the ``electric'' source contributions to the electromagnetic
fields.
\\

P. W. Bridgman\footnote{ P. W. Bridgman, 1961, 
{\it The Thermodynamics of Electrical Phenomena in Metals and a
Condensed Collection of Thermodynamic Formulas},
Dover Publications. -- for definitions of ``working'' and ``driving''
electromotive forces, see comments in preface page v, and text 
pages 19,61,63,69,129ff. And for comments relating to the expected, 
but as yet unseen, ``E.M.F. produced by temperature varying with
time,''  see preface page vi, and text pages 144-145.} observed, in
1961,  that thermoelectric phenomena
require the phenomenological description of e.m.f to allow for two
different kinds of electromotive force, one that provides what he
calls the ``working'' e.m.f, and the other that provides the
``driving'' e.m.f, for the thermoelectric system. The ``working''
e.m.f is responsible for the production of the total energy that
emerges from the system, while the ``driving'' e.m.f is responsible
for moving the charges in the system, giving rise to the electric
current. 

These two e.m.fs, traditionally considered the same normally
in electricity, are not the same when including thermoelectric
effects. 

Bridgman invents a thermodynamic construction to define
these two phenomenologically required e.m.fs, but he emphasises that 
since these are constructions they are not directly observable. Here
we find an alternative explanation of Bridgman's idea of the two
e.m.fs, on grounds much more fundamentally linked to the
electromagnetic equations and not requiring his ad hoc thermodynamic arguments. 
\\

If we take our new Coulomb-Gauss law, eqn (27),  and separate the sources so
that we isolate each effect in order to consider the impact of one 
independently of the other, we find the corresponding result to
Bridgman's two e.m.fs. 

We set the free electric charge density to
zero and label the electric field, whose source is given by the
temporal term, $+1/c.\partial T/\partial t$, alone, with a subscript to indicate this 
part arises from the temporal source; $\mathbf E_{T}$. 

\begin{equation}
div (\mathbf E_{T})  = +1/c.\partial T/\partial t
\end{equation}

Then, by subtracting this electric field from the total field we
obtain the Coulomb-Gauss law we are more familiar with, consisting of
just the free electric charge density as source. 
\begin{equation}
	div (\mathbf E - \mathbf E_{T})  = 4\pi \rho
\end{equation}

Here, we can define $\mathbf E$, as the ``driving'' field, since this
is clearly responsible for moving the charges through the region of
space. Then, $\mathbf E - \mathbf E_{T}$, is the ``working'' field,
that determines the total energy (i.e. net energy) delivered by the
field in the act of moving those charges---part of the energy that
would normally be observed due to a charge moving under the influence
of that total electric field being now reabsorbed instead by the
mechanics of the reversible thermoelectric heat in the internal
energy conversion process. 

It is not clear to me, however, whether 
these two e.m.fs should simply replace Bridgman's constructions, or 
whether Bridgman's e.m.fs should be considered to exist
independently, thus masking the presence of the quaternion results. 
\\

Bridgman also notes that symmetric thermoelectric arguments require
that a time varying temperature will induce a related thermoelectric 
electromotive force in the material undergoing the variation of
temperature, an effect that has not yet been experimentally observed.

Now, we would expect that the temporal field, T, in an homogeneous
material, would itself be uniform throughout the material medium, and
show no direct independent variation in space or time, except that, 
given the close relation to heat energy, the temporal field would
vary directly with temperature, K, and would consequently show an 
indirect space variation and time variation to the extent that the 
temperature itself varies with space and time.\footnote{We use the
letter,  K, for temperature, instead of the more usual letter, T, 
since the latter is being used here for the temporal field, and 
we can remember this as K for Kelvin instead of T for Thomson, 
given that William Thomson became Lord Kelvin, hence the promotion 
of the letter.}
\subsection{\label{sec:level2}dT/dK}
This relationship between the temporal field and temperature can be
more clearly expressed functionally, $T = T(K)$.
And with this in mind, we conclude that the time
rate of change of the temporal field is proportional to the time rate
of change of the temperature, 
\begin{equation}
      \frac{1}{c} \frac{\partial T}{\partial t} = \frac{1}{c}
      \frac{dT}{dK} \frac{\partial K}{\partial t}
\end{equation}

In fact, for any given homogeneous material medium, the measure of
the temporal field is simply a measure of the temperature, because
the parameter dT/dK is a characteristic of the medium. Given that we 
can write, $dT/dK = -1/(qc).d(-qTc)/dK$, and -qTc is the heat energy
absorbed per unit time by the charge, q, we see that dT/dK is
effectively the measure of a type of ``heat capacity'' of the unit charge
within the particular material.

Consequently, from the new Coulomb-Gauss law, we infer that an
electric field, $E_{T}$, is induced by a time rate of change in
temperature, the magnitude of the field being determined by 
this special ``heat capacity'' of a unit charge within the material and 
 the rate of change of temperature with time. The
direction of the field is radially away from the point at which the 
temperature change occurs. 

In an homogeneous isotropic medium, the 
neighbouring points also produce similar fields, but being oppositely
directed, they cancel each other, leaving no net field within the
medium. However, where there is an anisotropy introduced into the
medium, such as a temperature gradient within the material, a net 
field will emerge along the direction of that anisotropy. This is 
consistent with Bridgman's thermoelectric e.m.f induced by changing 
temperature with time, which requires the presence of an isothermal 
electric current, thus providing the requisite anisotropy, and 
also consistent with the pyroelectric effect where the mechanical 
stress and strain provide the anisotropic conditions. Thus, here
again, we find the effects of the temporal field being masked by 
other theoretically anticipated and experimentally known phenomena. 
\subsection{\label{sec:level2}Seebeck Effect}
Now, consider an electric charge moving from one material medium into
another, say two different metals. From the rest frame of the moving
charge, we find that there is a jump in the temporal field, T, at the
time the charge is seen to cross the boundary from one metal and
enter the other metal in the lab frame. This time rate of change of
temporal field is seen as an electric field, $E_{T}$, by the charge, 
according to our new Coulomb-Gauss law, and this produces an electric
force on the charge that accelerates it from rest in its
instantaneous rest frame, or just accelerates the charge in the lab
frame, and this is the source of the Seebeck thermoelectric e.m.f.

Note, no static surface charges are required at the boundary between
the two metals to establish an electrostatic field that will then 
accelerate the charges to sustain the current. The jump in the
temporal field between the two metals, which is essentially related
to the jump in the specific heat capacities per unit charge and the 
jump in electrical conductivity, is the sole source of the manifest
e.m.f. Randomly moving charges that cross the boundary will be
accelerated by this e.m.f, and a closed circuit will thus sustain a 
current. 

\subsection{\label{sec:level2}Thomson Effect}

Consider the new form of Ampere's law. If we multiply both sides of
the equation (25) by the ratio of current to electrical conductivity,
$\mathbf J/\sigma $, using the vector ``dot product,'' then write the gradient of 
the temporal field in terms of the temperature gradient, and
re-arrange the terms, we obtain four quantities that sum to zero,
\begin{eqnarray}
\frac{\mathbf J^2}{\sigma} & + & \frac{c}{4\pi \sigma} \frac{dT}{dK} \mathbf J\cdot grad(K) \\ \nonumber
      & - & \frac{c}{4\pi \sigma} \mathbf J \cdot curl(\mathbf B) 
 +  \frac{1}{4\pi \sigma} \mathbf J \cdot \partial \mathbf E/\partial t = 0. \\ \nonumber
\end{eqnarray}

The first term we recognise as the Joule Heat produced by an
electrical current.  It is proportional to the square of the
electrical current density, $\mathbf J^2$, and is thus independent of the
direction of the current or the sign of the charge carriers. 

The
second term we recognise has the form of the Thomson Heat produced by
an electrical current flowing up a temperature gradient. It is linear
in the electrical current density, $\mathbf J$, and thus reverses sign when the
current reverses direction or the signs of the charge carriers are
reversed.

These two terms must just balance the last two terms in this equation,
for a thermally isolated electromagnetic system. But if we design an
experiment to reduce the latter two terms to zero, say, for example, we arrange
things so that the electric field is constant in time, $\partial
\mathbf E/\partial t = 0$, and the circulation of the magnetic field
is perpendicular to the current flow, $\mathbf J \cdot curl(\mathbf
B) = 0$, then the system can no longer be thermally isolated, in
general, and we must either pump heat energy into or extract heat out
to maintain our particular requirements.

When placing this electromagnetic system in contact with heat
reservoirs, therefore, and allowing heat to be exchanged between
the system and the reservoirs, in such a way that the last two 
terms vanish, we obtain the conditions that characterize the 
Thomson Heat experiment. The rate of net heat energy exchanged
with the reserviors, $dQ/dt$, is now equal to the sum of the first two
heats, and the equation becomes the usual Thomson Heat equation
of thermoelectricity, which says the rate of production of heat by
the system is the sum of the irreversible Joule Heat and the
reversible Thomson Heat, 
\begin{eqnarray}
   \frac{dQ}{dt} & = & \mathbf J^2/\sigma - h_{T} \mathbf J\cdot grad(K)
   \ \ \ \ \ \ \ \ \\
   h_{T} & = & -\frac{c}{4\pi \sigma} \frac{dT}{dK} \ \ \ \ \ \ \ \ 
\end{eqnarray}

Where now, $h_{T}$, is the Thomson Specific Heat of the material,
which is defined to be the quantity of ``reversible heat'' absorbed
per unit time by an electrical current of unit strength flowing up a
temperature gradient of one degree per unit length in a wire of unit 
area cross section. The ``reversible
heat'' is separated from the ``irreversible heat'' due to Joule
heating by measuring the total heat produced by the current flowing
in one direction, then reversing the direction of the current and
measuring the total heat again. The difference between the two total 
heats is then twice the Thomson Heat. 
\\

We are now in a position to interpret the special ``heat capacity''
parameter, dT/dK, introduced above. It is essentially
equivalent to the product of the Thomson Specific Heat and the
Electrical Conductivity of the medium, $h_{T}\ \cdot \ \sigma$, and thus it is
indeed a characteristic of the material. 

\section{\label{sec:level1}POLARIZATION AND MAGNETIZATION.\protect\\}

We can identify the net electric field with the electric
displacement,  $\mathbf D = \mathbf E - \mathbf E_{T}$. Then, if we
define the magnetic field, $\mathbf B_{T}$, to be that which solves the
equation, $curl(\mathbf B_{T}) = +1/c.\partial \mathbf E_{T}/\partial t +
grad(T)$, we can write, $\mathbf H = \mathbf B - \mathbf B_{T}$, and
obtain the more familiar macroscopic equations.
\\

We can then argue that in a macroscopic medium, the effects due to
the presence of these new terms, $grad(T)$ and $+1/c.\partial
T/\partial t$, are normally
hidden in the complexities of the Polarization and Magnetization
parameters that describe the material, and thus, in many instances, the
effects of the temporal field are not easily separated to be
identified and measured as independent phenomena. 

\section{\label{sec:level1}CONCLUSION.\protect\\}

When James Clerk Maxwell\footnote{J. C. Maxwell, 1954, 
{\it Treatise on Electricity and Magnetism}, 3rd ed., 2 vols, Dover, New York.}
wrote the second edition of his {\it Treatise
on Electricity and Magnetism} he included a quaternion representation
of his electromagnetic equations, but he did not include both
left-hand and right-hand derivatives, and the differential operator
nabla was restricted to the 3-dimensional space form lacking a time
component, and so his work is
fundamentally different from that presented here. 
\\

Indeed, in the calculus of  quaternions the differential operator
almost always appears on the left acting towards the variable on the
right, ignoring the other alternative.
And even though, Charles Jasper Joly\footnote{C. J. Joly, 1905, 
{\it A Manual of Quaternions},  London Macmillan. In Art 57, Joly recognises the
two  different left and right differentiations; pp.74-77, and 
Exercises Ex.5, Ex.11, on pg.76.} notes the distinction in his
book  {\it A Manual of Quaternions}, the importance of the idea goes
unnoticed, unexplored, and unused. As a consequence of this, an 
important field component went missing
in Maxwell's Equations, and all of modern physics has developed from 
there perpetuating one of the consequences of this oversight, namely,
that the electromagnetic field possesses six components, whereas, as
we have shown, there should be seven. 
\\

Our macroscopic experience tells us that heat is produced by two
opposing agents acting, the one against the other, rubbing as it
were, as in the familiar case of mechanical friction, to produce the 
gross fire that manifests as heat when there is contact with matter. 
So, when we find the electric field is also the sum of two opposing 
principles, the left-handed tension acting against the right-handed 
tension, we should not be surprised to find a heat component, a more 
subtle fire, hiding within the field, and emerging as heat when the 
field comes into contact with charged matter. 

Indeed, the action of a material body can be generally described in 
terms of three transformations: translations, rotations, and
pulsations. The electric field, $\mathbf E$, tends to induce translations in a 
test charge, while the magnetic field, $\mathbf B$, tends to induce rotations
when the charge is moving, we are then left to conclude that the
temporal field, T, tends to induce the pulsations. 

Thus, with the inclusion of the missing temporal field the
description of the action of a charged material particle is
complete---we infer such a particle must have an extended structure
with a variable intrinsic pulse in addition to its quantum
mechanically determined fixed intrinsic spin. 
\\

This brief note introduces the essential ideas that will be explored
more fully in a future paper.
\\

\begin{acknowledgments}
I wish to acknowledge the encouragement from the usenet community of
sci.physics, sci.physics.particle, and sci.math, where the lively discussion my ideas
began in 1995, and to acknowledge  Prof. Pertti Lounesto in
particular, who kept prompting me to get these ideas published in a more formal medium.
\end{acknowledgments}

\end{document}